\documentclass[aps,prd,showpacs,showkeys,preprintnumbers,unsortedaddress,superscriptaddress,nofootinbib,notitlepage]{revtex4-1}
\usepackage{amsmath,mathrsfs,amssymb}
\usepackage{amsfonts}
\usepackage{amsmath}
\usepackage{hyperref}
\usepackage[english]{babel}
\usepackage{graphicx}
\usepackage{epsfig}
\usepackage{bm}
\usepackage{longtable}
\usepackage{verbatim}
\usepackage{longtable}

\newcommand{\mathsym}[1]{{}}

\input{tcilatex}
\begin{document}

\title{Viable textures for the fermion sector}
\date{\today}
\author{A. E. C\'arcamo Hern\'andez}
\email{antonio.carcamo@usm.cl}
\affiliation{{\small Universidad T\'ecnica Federico Santa Mar\'{\i}a and Centro Cient\'{\i}fico-Tecnol\'ogico de Valpara\'{\i}so}\\
Casilla 110-V, Valpara\'{\i}so, Chile}
\author{I. de Medeiros Varzielas}
\email{ivo.de@udo.edu}
\affiliation{{\small Department of Physics, University of Basel,}\\
Klingelbergstr. 82, CH-4056 Basel, Switzerland} 
\affiliation{{\small School of Physics and Astronomy, University of Southampton,}\\
Southampton, SO17 1BJ, U.K.}

\begin{abstract}
We consider a modification of the Fukuyama-Nishiura texture and compare it
to the precision quark flavour data, finding that it fits the data very well
but at the cost of accidental cancelations between parameters. We then
propose different viable textures for quarks, where only the Cabibbo mixing
arises from the down sector, and extend to the charged leptons while
constructing a complementary neutrino structure that leads to viable lepton
masses and mixing.
\end{abstract}

\maketitle

\section{Introduction}

The flavour puzzle is not understood in the context of the Standard Model
(SM), which does not specify the Yukawa structures and has no justification
for the number of generations. As such, extensions addressing the fermion
masses and mixing are particularly appealing.

In building models that address the flavour problem, it is important to know
structures that lead to the observed fermion flavour data. 
This approach was pioneered by \cite{Fritzsch:1977za}.
In this work
we introduce two proposals.

We start by revisiting the Fukuyama-Nishiura (FN) texture \cite%
{Fukuyama:1997ky,Matsuda,MatsudaRG}, which is no longer phenomenologically viable as
shown in \cite{Carcamo}. A simple modification is to modify the texture
slightly by enabling a non-zero 11 entry, which we show is already a viable
texture \footnote{%
Note added in proof: the same texture was very recently proposed, but for
the neutrinos only, in \cite{Gehrlein:2014wda}.}.

We then introduce other quark textures where the Cabibbo angle comes from
the down quarks whereas the other mixing angles come from the up sector,
which also successfully describes the quark masses and mixing. We extend
this to the lepton sector, with charged leptons sharing the texture of the
down quarks and the neutrinos significantly contributing to a viable PMNS
mixing matrix.

Other works in the literature considering textures include \cite{Xing,Peccei:1995fg,Roberts:2001zy}, and some recent works such as \cite{asgood,notasgood,uncompared}. A recent and more thorough review of textures can be found e.g. in \cite{Gupta:2012dma}.

\section{Textures for the quark sector}

\subsection{Fukuyama-Nishiura texture and its modification}

Proposed for leptons in \cite{Fukuyama:1997ky} and also used for quarks in 
\cite{Matsuda,MatsudaRG}, the FN texture consists in mass matrices of the form: 
\begin{equation}
\left( 
\begin{array}{ccc}
0 & A_{f} & A_{f} \\ 
A_{f} & B_{f} & C_{f} \\ 
A_{f} & C_{f} & B_{f}%
\end{array}%
\right) \allowbreak .
\end{equation}
This texture has only one zero and therefore falls outside the classes described in \cite{Gupta:2012dma}.
In \cite{Carcamo} it was shown that this texture doesn't quite work
currently, because it fails to reproduce the observed value of the CP
violating phase $\delta$. A simple modification of the texture is to
consider 
\begin{equation}
\widetilde{M}_{f}=\left( 
\begin{array}{ccc}
D_{f} & A_{f} & A_{f} \\ 
A_{f} & B_{f} & C_{f} \\ 
A_{f} & C_{f} & B_{f}%
\end{array}%
\right) \allowbreak ,
\end{equation}
with the complex phases included in the following way: 
\begin{equation}
M_{f}=P_{f}\widetilde{M}_{f}P_{f}^{\dagger },\quad \quad \quad \quad
P_{f}=\left( 
\begin{array}{ccc}
1 & 0 & 0 \\ 
0 & e^{-i\beta _{f}} & 0 \\ 
0 & 0 & e^{-i\gamma _{f}}%
\end{array}%
\right) ,
\end{equation}%
where $A_{f}$, $B_{f}$, $C_{f}$ and $D_{f}$ are real parameters. $\widetilde{%
M}_{f}$ is diagonalized by an orthogonal matrix $R_{f}$: 
\begin{eqnarray}
R_{f}^{T} \widetilde{M}_{f}R_{f} &=&diag\left(
-m_{f_{1}},m_{f_{2}},m_{f_{3}}\right) ,
\end{eqnarray}
\begin{equation}
R_{f}=\left( 
\begin{array}{ccc}
c_{f} & s_{f} & 0 \\ 
-\frac{s_{f}}{\sqrt{2}} & \frac{c_{f}}{\sqrt{2}} & -\frac{1}{\sqrt{2}} \\ 
-\frac{s_{f}}{\sqrt{2}} & \frac{c_{f}}{\sqrt{2}} & \frac{1}{\sqrt{2}}%
\end{array}%
\right) ,\quad \quad \quad \quad c_{f}=\sqrt{\frac{m_{f_{2}}-D_{f}}{%
m_{f_{2}}+m_{f_{1}}}},\quad \quad \quad \quad s_{f}=\sqrt{\frac{%
m_{f_{1}}+D_{f}}{m_{f_{2}}+m_{f_{1}}}},
\end{equation}%
with fermion masses given by:%
\begin{eqnarray}
-m_{f_{1}} &=&\frac{1}{2}\left( D_{f}+B_{f}+C_{f}-\sqrt{\left(
D_{f}-C_{f}-B_{f}\right) ^{2}+8A_{f}^{2}}\right) ,  \notag \\
m_{f_{2}} &=&\frac{1}{2}\left( D_{f}+B_{f}+C_{f}+\sqrt{\left(
D_{f}-C_{f}-B_{f}\right) ^{2}+8A_{f}^{2}}\right) ,  \notag \\
m_{f_{3}} &=&B_{f}-C_{f}.
\end{eqnarray}
while
\begin{widetext}
\begin{eqnarray}
V&=&O_{U}^{T}P_{UD}O_{D}=\left( 
\begin{array}{ccc}
c_{U}c_{D}+\frac{1}{2}s_{U}s_{D}\left( e^{i\sigma }+e^{i\tau }\right) & 
c_{U}s_{D}-\frac{1}{2}s_{U}c_{D}\left( e^{i\sigma }+e^{i\tau }\right) & 
\frac{1}{2}s_{U}\left( e^{i\sigma }-e^{i\tau }\right) \\ 
s_{U}c_{D}-\frac{1}{2}c_{U}s_{D}\left( e^{i\sigma }+e^{i\tau }\right) & 
s_{U}s_{D}+\frac{1}{2}c_{U}c_{D}\left( e^{i\sigma }+e^{i\tau }\right) & 
\frac{1}{2}c_{U}\left( e^{i\tau }-e^{i\sigma }\right) \\ 
\frac{1}{2}s_{D}\left( e^{i\sigma }-e^{i\tau }\right) & \frac{1}{2}%
c_{D}\left( e^{i\tau }-e^{i\sigma }\right) & \frac{1}{2}\left( e^{i\sigma
}+e^{i\tau }\right)%
\end{array}%
\right) \allowbreak
\end{eqnarray}
\end{widetext}
is the CKM quark mixing matrix,where $P_{UD}=P_{U}^{\dagger
}P_{D}=diag\left( 1,e^{i\sigma },e^{i\tau }\right) $, with $\sigma =\beta
_{U}-\beta _{D}$ and $\tau =\gamma _{U}-\gamma _{D}$.

While it is possible to fit our texture parameters to values extrapolated from the data up to high scales (such as the grand unified scale), as we are not considering any specific models we have no control over important details such as threshold corrections, which would affect the values of the masses and CKM entries we would fit to. The CKM entries in particular do not vary much, and CKM mixing angles which involve ratios between CKM entries almost do not vary with scale (see e.g. the analysis for the original FN texture in \cite{MatsudaRG} where the CKM entries vary by 15\% and the angles are nearly the scale independent).
We therefore choose to fit our texture parameters to the data at the $M_{Z}$ scale.
 We use the values of the quark masses at the $M_{Z}$ scale from \cite%
{Bora:2012tx} shown in Table \ref{Quarkmasses} (which are similar to those
in \cite{Xing:2007fb}), we constrained the parameters $A_{u,d}$, $B_{u,d}$
and $C_{u,d}$. We note that in order to get the hierarchy between the
masses, accidental cancelations between the parameters are required
(particularly in the up quark sector), but that this was already the case
for the original FN texture. %%%

We then varied the parameters $D_{u,d}$, $\sigma $ and $\tau $ we fitted the
magnitudes of the CKM matrix elements, the CP violating phase and the
Jarlskog invariant $J$ to the experimental values shown in Table \ref{Observables}. The experimental values of CKM magnitudes and Jarlskog invariant are taken
from \cite{PDG}. 

To find the values of the parameters that yield physical observables in the quark sector consistent with the experimental data, we minimize the following $\chi ^{2}$ function:
\begin{eqnarray}
\chi ^{2}\left( A_{u,d},B_{u,d},C_{u,d},D_{u,d},\sigma ,\tau \right)  &=&\sum
_{\substack{ i=u,c,t \\ j=d,s,b}}\left\vert \left( V_{ij}\right)
_{pred}-\left( V_{ij}\right) _{\exp }\right\vert ^{2}+\left( \frac{J_{pred}-J_{\exp
}}{J_{\exp }}\right) ^{2} %\\
%&&+\left\vert \left( V_{ud}\right) _{pred}-\left( V_{ud}\right) _{\exp}\right\vert ^{2}+\left\vert \left( V_{us}\right) _{pred}-\left(
%V_{us}\right) _{\exp }\right\vert ^{2}+\left\vert \left( V_{ub}\right)_{pred}-\left( V_{ub}\right) _{\exp }\right\vert ^{2} \\
%&&+\left\vert \left( V_{cd}\right) _{pred}-\left( V_{cd}\right) _{\exp}\right\vert ^{2}+\left\vert \left( V_{cs}\right) _{pred}-\left(V_{cs}\right) _{\exp }\right\vert ^{2}+\left\vert \left( V_{cb}\right)
%_{pred}-\left( V_{cb}\right) _{\exp }\right\vert ^{2} \\
%&&+\left\vert \left( V_{td}\right) _{pred}-\left( V_{td}\right) _{\exp}\right\vert ^{2}+\left\vert \left( V_{ts}\right) _{pred}-\left(V_{ts}\right) _{\exp }\right\vert ^{2}+\left\vert \left( V_{tb}\right)_{pred}-\left( V_{tb}\right) _{\exp }\right\vert ^{2}
\label{chifunction}
\end{eqnarray}

\begin{table}[tbh]
\begin{center}
\begin{tabular}{c|l}
\hline\hline
Quark masses & Experimental Value \\ \hline
$m_{d}(MeV)$ & $2.9_{-0.4}^{+0.5}$ \\ \hline
$m_{s}(MeV)$ & $57.7_{-15.7}^{+16.8}$ \\ \hline
$m_{b}(MeV)$ & $2820_{-40}^{+90}$ \\ \hline
$m_{u}(MeV)$ & $1.45_{-0.45}^{+0.56}$ \\ \hline
$m_{c}(MeV)$ & $635\pm 86$ \\ \hline
$m_{t}(GeV)$ & $172.1\pm 0.6\pm 0.9$ \\ \hline\hline
\end{tabular}%
\end{center}
\caption{Quark masses at the $M_{Z}$ scale \protect\cite{Bora:2012tx}.}
\label{Quarkmasses}
\end{table}
\begin{table}[tbh]
\begin{center}
\begin{tabular}{c|l|l}
\hline\hline
Observable & Modified FN texture & Experimental Value \\ \hline
$\bigl|V_{ud}\bigr|$ & \quad $0.974$ & \quad $0.97427\pm 0.00015$ \\ \hline
$\bigl|V_{us}\bigr|$ & \quad $0.225$ & \quad $0.22534\pm 0.00065$ \\ \hline
$\bigl|V_{ub}\bigr|$ & \quad $0.00351$ & \quad $%
0.00351_{-0.00014}^{+0.00015} $ \\ \hline
$\bigl|V_{cd}\bigr|$ & \quad $0.225$ & \quad $0.22520\pm 0.00065$ \\ \hline
$\bigl|V_{cs}\bigr|$ & \quad $0.973$ & \quad $0.97344\pm 0.00016$ \\ \hline
$\bigl|V_{cb}\bigr|$ & \quad $0.0412$ & \quad $0.0412_{-0.0005}^{+0.0011}$
\\ \hline
$\bigl|V_{td}\bigr|$ & \quad $0.00867$ & \quad $%
0.00867_{-0.00031}^{+0.00029} $ \\ \hline
$\bigl|V_{ts}\bigr|$ & \quad $0.0404$ & \quad $0.0404_{-0.0005}^{+0.0011}$
\\ \hline
$\bigl|V_{tb}\bigr|$ & \quad $0.999$ & \quad $%
0.999146_{-0.000046}^{+0.000021}$ \\ \hline
$J$ & \quad $2.96\times 10^{-5}$ & \quad $(2.96_{-0.16}^{+0.20})\times
10^{-5}$ \\ \hline
$\delta $ & \quad $69.2^{\circ }$ & \quad $68^{\circ }$ \\ \hline\hline
\end{tabular}%
\end{center}
\caption{Comparison of our fit for the modified FN texture to the
experimental CKM magnitudes, CP violating phase and Jarlskog invariant.}
\label{Observables}
\end{table}

Our results, for which the $\chi ^{2}\left(A_{u,d},B_{u,d},C_{u,d},D_{u,d},\sigma
,\tau \right) $ function takes its minimum value $3.8\times 10^{-7}$, correspond to the following parameters: 
\begin{eqnarray}
A_{u} &=&3.80\times 10^{-2}GeV,\quad \quad \quad \quad B_{u}=86.37GeV,\quad
\quad \quad \quad C_{u}=-85.73GeV,\quad \quad \quad \quad   \notag \\
D_{u} &=&3.13\times 10^{-3}GeV,\quad \quad \quad \quad \sigma =87.9^{\circ },
\notag \\
A_{d} &=&8.79\times 10^{-3}GeV,\quad \quad \quad \quad B_{d}=1.44GeV,\quad
\quad \quad \quad C_{d}=-1.38GeV,  \notag \\
D_{d} &=&-2.35\times 10^{-4}GeV,\quad \quad \quad \quad \tau =92.6^{\circ }.
\end{eqnarray}%
The obtained magnitudes of the CKM matrix elements, the CP violating phase
and the Jarlskog invariant are in excellent agreement with the experimental
data. The required cancelation which we mentioned before is apparent e.g. in the values for $B_f$ and $C_f$.

\subsection{Mixing inspired textures}

Given that the modified FN textures lead to very good fits but suffer the
problem of requiring accidental cancelations, it is useful to obtain
alternative viable textures that have a similar $M_f M_f^\dagger$, and would
therefore also reproduce the CKM mixing well (note that as $M_f$ is
symmetric in both the original and modified FN textures, $M_f$ and $M_f
M_f^\dagger$ share the same structure). %%%
For this reason we consider now the following Mixing Inspired (MI) textures,
where we consider non-zero off-diagonal entries leading to the Cabibbo angle
in the down sector, and for the ups leading to the remaining (small) mixing
angles: 
\begin{eqnarray}
M_{U}&=&\frac{v}{\sqrt{2}}\left( 
\begin{array}{ccc}
c_{1}\lambda ^{8} & 0 & a_{1}\lambda ^{3} \\ 
0 & b_{1}\lambda ^{4} & a_{2}\lambda ^{2} \\ 
0 & 0 & a_{3}%
\end{array}%
\right) ,  \notag \\
M_{D}&=&\frac{v}{\sqrt{2}}\left( 
\begin{array}{ccc}
e_{1}\lambda ^{7} & f_{1}\lambda ^{6} & 0 \\ 
0 & f_{2}\lambda ^{5} & 0 \\ 
0 & 0 & g_{1}\lambda ^{3}%
\end{array}%
\right) ,  \label{Quarktextures}
\end{eqnarray}
where $\lambda =0.225$ is one of the Wolfenstein parameters, $v=246$ GeV\
the symmetry breaking scale and $a_{k}$ ($k=1,2,3$), $b_{1}$, $c_{1}$, $%
g_{1} $, $f_{1}$, $f_{2}$, $e_{1}$ are $\mathcal{O}(1)$ parameters.
From
comparison with the Wolfenstein parameterisation, we conclude that most
dimensionless parameters given in Eq. (\ref{Quarktextures}) can be real,
excepting $a_{1}$.

These $M_f$ matrices are not Hermitian, and therefore fall outside the classes considered in \cite{Gupta:2012dma}.
Furthermore, as we have discussed above, we have obtained these (non-symmetric) $M_f$ matrices from attempting a specific generalisation of the modified FN textures discussed in the previous section. Although the result turns out to be somewhat similar to the (symmetric) mass matrices that \cite{Peccei:1995fg} refers to as natural mass matrices, our MI textures are nevertheless distinct from the textures considered therein.

In order to obtain the quark masses and CKM mixing we diagonalize the Hermitian quantities $M_U M_U^\dagger$ and $M_D M_D^\dagger$.
Therefore, the up and down type quark masses are approximately given by: 
\begin{eqnarray}
m_{u} &\simeq &c_{1}\lambda ^{8}\frac{v}{\sqrt{2}},\quad \quad \quad
m_{c}\simeq b_{1}\lambda ^{4}\frac{v}{\sqrt{2}},\quad \quad \quad
m_{t}\simeq a_{3}\frac{v}{\sqrt{2}},  \notag \\
m_{d} &\simeq &\frac{e_{1}\lambda ^{7}}{\sqrt{2}}v,\quad \quad \quad
m_{s}\simeq f_{2}\lambda ^{5}\frac{v}{\sqrt{2}},\quad \quad \quad
m_{b}\simeq g_{1}\lambda ^{3}\frac{v}{\sqrt{2}}.  \label{quarkmasses}
\end{eqnarray}
It is noteworthy that Eq. (\ref{quarkmasses}) provide an elegant
understanding of all SM quark masses in terms of the Wolfenstein parameter $%
\lambda =0.225$ and of parameters of order unity. Note that all physical
parameters in the quark sector are linked with the electroweak symmetry
breaking scale $v=246$ GeV through their scalings by powers of the
Wolfenstein parameter $\lambda =0.225$, with $\mathcal{O}(1)$ coefficients.

The Wolfenstein parameterisation \cite{Wolfenstein:1983yz} of the CKM matrix
is: 
\begin{equation}
V_{W}\simeq \left( 
\begin{array}{ccc}
1-\frac{\lambda ^{2}}{2} & \lambda & A\lambda ^{3}(\rho -i\eta ) \\ 
-\lambda & 1-\frac{\lambda ^{2}}{2} & A\lambda ^{2} \\ 
A\lambda ^{3}(1-\rho -i\eta ) & -A\lambda ^{2} & 1%
\end{array}%
\right) ,  \label{wolf}
\end{equation}%
with 
\begin{eqnarray}
\lambda &=&0.22535\pm 0.00065,\quad \quad \quad A=0.811_{-0.012}^{+0.022}, \\
\quad \overline{{\rho }} &=&0.131_{-0.013}^{+0.026},\quad \quad \quad 
\overline{{\eta }}=0.345_{-0.014}^{+0.013}, \\
\overline{{\rho }} &\simeq &\rho \left( 1-\frac{{\lambda }^{2}}{2}\right)
,\quad \quad \quad \overline{{\eta }}\simeq \eta \left( 1-\frac{{\lambda }%
^{2}}{2}\right) .
\end{eqnarray}%
From the comparison with (\ref{wolf}), we find: 
\begin{eqnarray}
a_{3} &\simeq &1,\quad \quad \quad a_{2}\simeq A,\quad \quad \quad
a_{1}\simeq -A\sqrt{\rho ^{2}+\eta ^{2}}e^{i\delta }, \\
b_{1} &\simeq &\frac{m_{c}}{\lambda ^{4}m_{t}}\simeq 1.43,\quad \quad \quad
c_{1}\simeq \frac{m_{u}}{\lambda ^{8}m_{t}}\simeq 1.27,
\end{eqnarray}%
note that $a_{1}$ is required to be complex.

We fit the parameters $e_{1}$, $f_{1}$, $f_{2}$ and $g_{1}$ in Eq. (\ref%
{Quarktextures}) to reproduce the down type quark masses and quark mixing
parameters. The results for the CKM matrix elements, the Jarlskog invariant $%
J$ and the CP violating phase $\delta $ in Tables \ref{Observables0} and \ref%
{Observables2} correspond to the best fit values: 
\begin{equation}
e_{1}\simeq 0.6,\quad \quad \quad f_{1}\simeq 0.59,\quad \quad \quad
f_{2}\simeq 0.57,\quad \quad \quad g_{1}\simeq 1.42.
\end{equation}%
The experimental values of the CKM magnitudes and the Jarlskog invariant are
taken from Ref. \cite{PDG}. %%%%
\begin{table}[tbh]
\begin{center}
\begin{tabular}{c|l|l}
\hline\hline
Observable & MI textures & Experimental Value \\ \hline
$m_{u}(MeV)$ & \quad $1.47$ & \quad $1.45_{-0.45}^{+0.56}$ \\ \hline
$m_{c}(MeV)$ & \quad $641$ & \quad $635\pm 86$ \\ \hline
$m_{t}(GeV)$ & \quad $172.2$ & \quad $172.1\pm 0.6\pm 0.9$ \\ \hline
$m_{d}(MeV)$ & \quad $3.00$ & \quad $2.9_{-0.4}^{+0.5}$ \\ \hline
$m_{s}(MeV)$ & \quad $59.2$ & \quad $57.7_{-15.7}^{+16.8}$ \\ \hline
$m_{b}(GeV)$ & \quad $2.82$ & \quad $2.82_{-0.04}^{+0.09}$ \\ \hline
$m_{e}(MeV)$ & \quad $0.487$ & \quad $0.487$ \\ \hline
$m_{\mu }(MeV)$ & \quad $102.8$ & \quad $102.8\pm 0.0003$ \\ \hline
$m_{\tau }(GeV)$ & \quad $1.75$ & \quad $1.75\pm 0.0003$ \\ \hline
\end{tabular}%
\end{center}
\caption{MI textures and experimental values of charged fermion masses.}
\label{Observables0}
\end{table}
\begin{table}[tbh]
\begin{center}
\begin{tabular}{c|l|l|l}
\hline\hline
Obs. & Wolfenstein & MI textures & Experimental \\ \hline
$\bigl|V_{ud}\bigr|$ & \quad $0.9746$ & \quad $0.9743$ & \quad $0.97427\pm
0.00015$ \\ \hline
$\bigl|V_{us}\bigr|$ & \quad $0.22535$ & \quad $0.2253$ & \quad $0.22534\pm
0.00065$ \\ \hline
$\bigl|V_{ub}\bigr|$ & \quad $0.00868$ & \quad $0.00351$ & \quad $%
0.00351_{-0.00014}^{+0.00015}$ \\ \hline
$\bigl|V_{cd}\bigr|$ & \quad $0.22535$ & \quad $0.22501$ & \quad $0.22520\pm
0.00065$ \\ \hline
$\bigl|V_{cs}\bigr|$ & \quad $0.9746$ & \quad $0.97349$ & \quad $0.97344\pm
0.00016$ \\ \hline
$\bigl|V_{cb}\bigr|$ & \quad $0.0412$ & \quad $0.0411$ & \quad $%
0.0412_{-0.0005}^{+0.0011}$ \\ \hline
$\bigl|V_{td}\bigr|$ & \quad $0.00342$ & \quad $0.0110$ & \quad $%
0.00867_{-0.00031}^{+0.00029}$ \\ \hline
$\bigl|V_{ts}\bigr|$ & \quad $0.0412$ & \quad $0.0398$ & \quad $%
0.0404_{-0.0005}^{+0.0011}$ \\ \hline
$\bigl|V_{tb}\bigr|$ & \quad $1$ & \quad $0.999147$ & \quad $%
0.999146_{-0.000046}^{+0.000021}$ \\ \hline
$J$ & \quad $2.90\times 10^{-5}$ & \quad $2.94\times 10^{-5}$ & \quad $%
(2.96_{-0.16}^{+0.20})\times 10^{-5}$ \\ \hline
$\delta $ & \quad $69^{\circ }$ & \quad $68^{\circ }$ & \quad $68^{\circ }$
\\ \hline\hline
\end{tabular}%
\end{center}
\caption{Wolfenstein, MI textures and experimental values of CKM parameters.}
\label{Observables2}
\end{table}
As can be seen, the quark masses and the CKM matrix obtained from these
textures are in excellent agreement with the experimental data. The
agreement with the experimental data is as good as in the models of Refs. 
\cite{asgood} and better than, for example, those in Refs.~\cite{notasgood},

Given the success of these textures, it would be interesting to obtain them
as the consequence of a flavour symmetry model which could explain the
differing powers of $\lambda$ and the texture zeros. Nonetheless, the
derivation of these textures in the context of a flavour model is beyond the
scope of this work. %%%

\section{Lepton sector}

Within the context of a flavour symmetry model, if the charged leptons have
similar assignments as the quarks, it would be natural that they share also
a similar texture. In order to check if such an idea is viable, %%%
we extend the texture we used for the down quarks to the charged leptons: 
\begin{equation}
M_{l}=\frac{v}{\sqrt{2}}\left( 
\begin{array}{ccc}
x_{1}\lambda ^{8} & y_{1}\lambda ^{6} & 0 \\ 
0 & y_{2}\lambda ^{5} & 0 \\ 
0 & 0 & z_{3}\lambda ^{3}%
\end{array}%
\right) .
\end{equation}
Therefore, $M_{l}M_{l}^{\dagger}$ can be approximately diagonalized by a rotation
matrix $R_{l}$ according to: 
\begin{equation}
R_{l}^{T}M_{l}M_{l}^{\dagger}R_{l}\simeq \left( 
\begin{array}{ccc}
m_{e}^{2} & 0 & 0 \\ 
0 & m_{\mu }^{2} & 0 \\ 
0 & 0 & m_{\tau }^{2}%
\end{array}%
\right) ,\,\quad \quad \quad R_{l}\simeq \left( 
\begin{array}{ccc}
\cos \gamma & \sin \gamma & 0 \\ 
-\sin \gamma & \cos \gamma & 0 \\ 
0 & 0 & 1%
\end{array}%
\right) \allowbreak .  \notag
\end{equation}

The charged lepton masses are approximately: 
\begin{equation}
m_{e}\simeq x_{1}\lambda ^{8}\frac{v}{\sqrt{2}},\,\quad \quad \quad m_{\mu
}\simeq y_{2}\lambda ^{5}\frac{v}{\sqrt{2}},\,\quad \quad \quad m_{\tau
}\simeq z_{3}\lambda ^{3}\frac{v}{\sqrt{2}}.  \label{chargedleptonmasses}
\end{equation}

Note the remarkable feature that charged lepton masses are linked with the
scale of electroweak symmetry breaking $v=246$ GeV through their power
dependence on the Wolfenstein parameter $\lambda =0.225$, with $\mathcal{O}%
(1)$ coefficients.

In order to extend the MI textures to the lepton sector we also consider a
neutrino structure, within a type I seesaw implementation, with a $6\times 6$
neutrino mass matrix: 
\begin{equation}
M_{\nu }=\left( 
\begin{array}{cc}
0_{3\times 3} & M_{\nu }^{D} \\ 
\left( M_{\nu }^{D}\right) ^{T} & M_{R}%
\end{array}%
\right) ,
\end{equation}
\begin{eqnarray}
M_{\nu }^{D}&=\left( 
\begin{array}{ccc}
0 & \lambda \varepsilon _{12}^{\left( \nu \right) }\frac{v}{\sqrt{2}} & 0 \\ 
\lambda \varepsilon _{21}^{\left( \nu \right) }\frac{v}{\sqrt{2}} & 0 & 0 \\ 
\lambda \varepsilon _{31}^{\left( \nu \right) }\frac{v}{\sqrt{2}} & \lambda
^{2}\varepsilon _{32}^{\left( \nu \right) }\frac{v}{\sqrt{2}} & 0%
\end{array}%
\right) =\left( 
\begin{array}{ccc}
0 & c & 0 \\ 
a & 0 & 0 \\ 
b & d & 0%
\end{array}%
\right) ,  \notag \\
M_{R}&=\left( 
\begin{array}{ccc}
M_{1} & 0 & 0 \\ 
0 & M_{2} & 0 \\ 
0 & 0 & M_{3}%
\end{array}%
\right) =\left( 
\begin{array}{ccc}
w^{-1} & 0 & 0 \\ 
0 & p^{-1} & 0 \\ 
0 & 0 & q^{-1}%
\end{array}%
\right).
\end{eqnarray}

With $\left( M_{R}\right) _{ii}>>v$, the light neutrino mass matrix is given
by the type I seesaw formula:

\begin{equation}
M_{L}=M_{\nu }^{D}M_{R}^{-1}\left( M_{\nu }^{D}\right) ^{T}=\left( 
\begin{array}{ccc}
c^{2}p & 0 & cdp \\ 
0 & a^{2}w & abw \\ 
cdp & abw & wb^{2}+pd^{2}%
\end{array}%
\right) =\left( 
\begin{array}{ccc}
D & 0 & C \\ 
0 & A & B \\ 
C & B & \frac{B^{2}}{A}+\frac{C^{2}}{D}%
\end{array}%
\right) .
\end{equation}

Varying the parameters $A$, $B$, $C$, $D$\ and $\gamma $ we fitted $\Delta
m_{21}^{2}$, $\Delta m_{31}^{2}$\ (note that we define $\Delta
m_{ij}^{2}=m_{i}^{2}-m_{j}^{2}$), $\sin ^{2}\theta _{12}$, $\sin ^{2}\theta
_{13}$ and $\sin ^{2}\theta _{23}$\ to the experimental values \cite%
{Tortola:2012te} in Table \ref{NH} for the normal hierarchy neutrino mass
spectrum. The best fit result is: 
\begin{eqnarray}
\Delta m_{21}^{2} &=&7.62\times 10^{-5}eV^{2},\quad \quad \quad \Delta
m_{31}^{2}=2.55\times 10^{-3}eV^{2},  \notag \\
m_{\nu _{1}} &=&0,\,\quad \quad \quad m_{\nu _{2}}\simeq 9meV,\quad \quad
\quad \,m_{\nu _{3}}\simeq 50meV,\,\quad \quad \quad \gamma \simeq -0.035\pi
,  \notag \\
\sin ^{2}\theta _{12} &=&0.32,\quad \quad \quad \sin ^{2}\theta
_{13}=0.0246,\quad \quad \quad \sin ^{2}\theta _{23}=0.613,  \notag \\
A &\simeq &3.35\times 10^{-2}eV,\quad \quad \quad B\simeq 2.17\times
10^{-2}eV,  \notag \\
C &\simeq &5.38\times 10^{-3}eV,\quad \quad \quad \,D\simeq 3.60\times
10^{-3}eV.  \label{ParameterfitNH}
\end{eqnarray}

Comparing Eq (\ref{ParameterfitNH}) with Table \ref{NH} we see that the mass
squared splittings $\Delta m_{21}^{2}$ and $\Delta m_{31}^{2}$ and mixing
parameters $\sin ^{2}\theta _{12}$, $\sin ^{2}\theta _{13}$ and $\sin
^{2}\theta _{23}$ are in excellent agreement with the experimental data.
Note that here we considered all leptonic parameters to be real for
simplicity, but a non-vanishing CP phase in the PMNS mixing matrix can be
generated by making e.g. $y_1$ or $y_2$ complex.

\begin{widetext}

\begin{table}[tbh]
\begin{tabular}{|c|c|c|c|c|c|}
\hline
Parameter & $\Delta m_{21}^{2}$($10^{-5}$eV$^2$) & $\Delta m_{31}^{2}$($%
10^{-3}$eV$^2$) & $\left( \sin ^{2}\theta _{12}\right) _{\exp }$ & $\left(
\sin ^{2}\theta _{23}\right) _{\exp }$ & $\left( \sin ^{2}\theta
_{13}\right) _{\exp }$ \\ \hline
Best fit & $7.62$ & $2.55$ & $0.320$ & $0.613$ & $0.0246$ \\ \hline
$1\sigma $ range & $7.43-7.81$ & $2.46-2.61$ & $0.303-0.336$ & $0.573-0.635$
& $0.0218-0.0275$ \\ \hline
$2\sigma $ range & $7.27-8.01$ & $2.38-2.68$ & $0.29-0.35$ & $0.38-0.66$ & $%
0.019-0.030$ \\ \hline
$3\sigma $ range & $7.12-8.20$ & $2.31-2.74$ & $0.27-0.37$ & $0.36-0.68$ & 
\\ \hline
\end{tabular}%
\caption{Experimental ranges of leptonic mixing parameters from \cite{Tortola:2012te} for the
case of normal hierarchy.}
\label{NH}
\end{table}
\end{widetext}

\section{Conclusions}

In this paper we studied two sets of textures. We started with a simple
modification of the Fukuyama-Nishiura texture and showed that the extra
parameter makes it viable in the quark sector. The CKM mixing is fitted very
well, but the quark mass hierarchy can only be obtained through accidental
cancelations of the parameters (a problem it shares with the original
Fukuyama-Nishiura texture). With the aim of preserving the overall structure
that led to the good fit to CKM mixing, we then considered mixing inspired
textures based on the idea that the Cabibbo mixing originates from the down
sector whereas the up sector contributes to the remaining mixing angles.
From the mixing inspired textures we are able to obtain a very good fit to
the quark masses and CKM mixing without requiring accidental cancelations of the texture parameters. Finally we extended the mixing
inspired texture to the leptons: by reusing the down quark texture in the
charged lepton sector, and proposing a compatible texture for the neutrinos,
such that we obtain the observed leptonic mass and mixing parameters. As
such these textures represent a good starting point for a flavour model
(under which the charged leptons and down quarks could transform similarly). 
%%%

\section*{Acknowledgments}

This project has received funding from the Swiss National Science
Foundation. This project has received funding from the European Union's
Seventh Framework Programme for research, technological development and
demonstration under grant agreement no PIEF-GA-2012-327195 SIFT. This
project was also supported by Fondecyt (Chile), Grant No. 11130115 and by
DGIP internal Grant No. 111458. A.E.C.H thanks University of Basel for
hospitality and for partially financing his visits. The visits of AECH to
the University of Basel were also financed by Fondecyt (Chile), Grant No.
11130115 and by DGIP internal Grant No. 111458.

\end{document}